\documentclass[12pt]{article}
\newcommand{\bd}[1]{ \mbox{\boldmath $#1$} }
\usepackage{psfig}
\begin{document}
\def\ii{\'\i}

\title{A schematic model for QCD at finite temperature.}

\author{
S. Lerma H.\thanks{e-mail: alerma@nuclecu.unam.mx},
S. Jesgarz \thanks{e-mail: jesgarz@nuclecu.unam.mx},
P. O. Hess$^1$ \thanks{e-mail: hess@nuclecu.unam.mx}, \\
O. Civitarese$^2$ \thanks{e-mail: civitare@venus.fisica.unlp.edu.ar},
and M. Reboiro$^2$ \thanks{e-mail: reboiro@venus.fisica.unlp.edu.ar},
\\
\\ {\small\it$^{1}$Instituto de Ciencias Nucleares, Universidad
Nacional Aut\'onoma de M\'exico,} \\
{\small\it Apdo. Postal 70-543, M\'exico 04510 D.F.} \\
{\small\it$^{2}$ Departamento de F\'{\i}sica, Universidad Nacional de La Plata, } \\
{\small\it c.c. 67 1900, La Plata, Argentina. } \\
$\;\;\;\;\;\;\;\;\;\;$\\ }
\maketitle

\noindent {\small {\bf Abstract}: The simplest version of a class
of toy models for QCD is presented. It is a Lipkin-type model, for
the quark-antiquark sector, and, for the gluon sector, gluon pairs
with spin zero are treated as elementary bosons. The model
restricts to mesons with spin zero and to few baryonic states. The
corresponding energy spectrum is discussed. We show that ground
state correlations are essential to describe physical properties
of the spectrum at low energies. Phase transitions are described
in an effective manner, by using coherent states. The appearance
of a Goldstone boson for large values of the interaction strength
is discussed, as related to a collective state. The formalism is
extended to consider finite temperatures. The partition function
is calculated, in an approximate way, showing the convenience of
the use of coherent states. The energy density, heat capacity and
transitions from the hadronic phase to the quark-gluon plasma are
calculated.
} \\


\section{Introduction}

QCD is considered to be the theory of the strong interactions. It
is well understood at high energies. At low energies, the QCD
coupling constant becomes too large to apply perturbation theory.
Lattice gauge calculations \cite{latticebook} may describe the
non-perturbative QCD regime, instead. Yet, problems like finite
size effects \cite{lattice} and fermion doubling still persist.
Although some advances have been made, only the lowest states for
a given spin and parity can be calculated. The observed sequence
of levels cannot be explained by lattice gauge calculations and
alternative methods have to be develop to explain the ordering
\cite{gluons99}. Many effective models have achieved some success
in describing the low energy regime of QCD \cite{swan,bonn}. These
models have in common that only quarks and antiquarks are taken
into account in the fermionic sector, while effective gluon
potentials or states with a fix number of gluons are considered.
In the real world hadrons are built by quarks, antiquarks and
gluons \cite{mixing}. The interactions between these degrees of
freedom, in consequence, may play an essential role in order to
understand QCD at low energies.

Concerning QCD effects at finite temperature, i.e. the
investigation of the quark-gluon plasma (QGP) \cite{qgp}, it
exists an intense effort, mainly focussed on computational aspects
of the problem \cite{qmd}.

In this paper, we propose a toy model for QCD which: i) is
amenable for an analytical treatment, except for a numerical
matrix diagonalization, ii) may describe the meson spectrum for
flavor $(0,0)$-spin 0 and partly the baryon spectrum, iii) it is
able to describe phase transitions at $T=0$, as a function of
coupling parameters, iv) can describe some characteristics of the
transition from the hadron gas to the QGP , and v) can be used to
test microscopic many-body techniques, intended to describe
realistic scenarios of QCD which cannot be accessed by other
methods.

The model is meant to mock up the basis features of
non-perturbative QCD, in a similar way as some schematic models do
in the nuclear many-body problem \cite{lipkin,schutte}. We should
stress that our toy model does not result from a field theory,
rather it represents the interactions between effective degrees of
freedom, instead. The basic ideas, procedures and methods can be
discussed already for the simplest versions of the toy model in a
very transparent way. At a certain point in our discussion we
shall concentrate on the structure of an extended version of the
model, which can be treated analytically.

The paper is organized as follows. In section 2 the model is
introduced, the energy spectrum is calculated and its structure is
discussed in terms of the elementary degrees of freedom. Coherent
states are introduced to determine the occurrence of phase
transition, induced by variations of the strength of the
interactions. In section 3 we discuss finite temperature effects,
by introducing temperature and by calculating the grand canonical
partition function. Conclusions are drawn in section 4.

\section{The Toy Model: Zero Temperature Case.}

The fermion sector of the theory is described by the action of the
operators which create (annihilate) quarks, with effective masses
$\omega_f$. Schematically, it corresponds to the situation
represented in Figure 1, where two levels, with energy $\pm
\omega_f$ and degeneracy $2\Omega$ are represented as valence
space \cite{lang}. The degeneracy of each level, is given by the
product of the number of colors ($n_c$), spin ($n_S$), flavors
 ($n_{fl}$) and all other possible degrees of freedom ($n_{col}$), like
orbital quantum numbers, etc.. For temperature $T=0$ and no
interaction the lower level is filled by fermions. The creation
(annihilation) operators of these fermions are
$\bd{c}^{\dagger}_{\alpha (1,0)f \sigma i}$ ($\bd{c}^{\alpha
(1,0)f \sigma i}$), in co- and contra-variant notation for the
indices. The symbol $(1,0)f$ refers to the flavor part, where
$(1,0)$ is the SU(3)-flavor notation and $f$ is a short hand
notation for the hypercharge $Y$, the isospin $T$ and its third
component $T_z$. The index $\sigma$ represents the two spin
components $\pm \frac{1}{2}$, the index $i$ $=$ 1 or 2, stands for
the upper or lower level and the index $\alpha$ represents all
remaining degrees of freedom, which are at least 3 if the color
degree of freedom is considered. Lowering and raising the indices
of the operators introduces a phase, which depends on the
convention used \cite{draayer}, and a change of the indices to
their conjugate values, i.e., the quantum numbers $(1,0)YTT_z
\sigma$ change to $(0,1)-YT-T_z-\sigma$.

The quark and antiquark creation and annihilation operators are
given in terms of the operators $c$ and $c^{\dagger}$

\begin{eqnarray}
\bd{a}^\dagger_{\alpha f \sigma} & = & \bd{c}^\dagger_{\alpha f \sigma 1}, ~~~
\bd{d}_{\alpha f \sigma}  =  \bd{c}^\dagger_{\alpha f \sigma 2}
\nonumber \\
\bd{a}^{\alpha f \sigma} & = & \bd{c}^{\alpha f \sigma 1}, ~~~
\bd{d}^{\dagger ~\alpha f \sigma}  =  \bd{c}^{\alpha f \sigma 2}
~~~,
\label{c}
\end{eqnarray}
which corresponds to the Dirac picture of particles and
antiparticles: quarks are described by fermions in the upper level
and antiquarks by holes in the lower level.

The gluon sector of the model space is described by bosons which
represent pair of gluons coupled to spin zero. The energy of a
boson state is fixed at the value $\omega_b$ and the state is
created (annihilated) by the action of a boson creation
(annihilation) operator $\bd{b}$ on the vacuum.

The quark-antiquark pairs of the model are given by

\begin{eqnarray}
\bd{C}_{f_1 \sigma_1 1}^{f_2 \sigma_2 2} & = & \bd{B}_{f_1 \sigma_1}^{\dagger f_2 \sigma_2}
= \sum_\alpha \bd{c}^\dagger_{\alpha f_1 \sigma_1 1} \bd{c}^{\alpha f_2 \sigma_2 2}
= \sum_\alpha \bd{a}^\dagger_{\alpha f_1 \sigma_1} \bd{d}^{\dagger \alpha f_2 \sigma_2}
\nonumber \\
\bd{C}_{f_1 \sigma_1 2}^{f_2 \sigma_2 1} & = & \bd{B}_{f_1 \sigma_1}^{f_2 \sigma_2}
= \sum_\alpha \bd{c}^\dagger_{\alpha f_1 \sigma_1 2} \bd{c}^{\alpha f_2 \sigma_2 1}
= \sum_\alpha \bd{d}_{\alpha f_1 \sigma_1} \bd{a}^{\alpha f_2 \sigma_2}
\nonumber \\
\bd{C}_{f_1 \sigma_1 1}^{f_2 \sigma_2 1} & = &
\sum_\alpha \bd{c}^\dagger_{\alpha f_1 \sigma_1 1} \bd{c}^{\alpha f_2 \sigma_2 1} =
\sum_\alpha
\bd{a}^\dagger_{\alpha f_1 \sigma_1} \bd{a}^{\alpha f_2 \sigma_2} \nonumber \\
\bd{C}_{f_1 \sigma_1 2}^{f_2 \sigma_2 2} & = &
\sum_\alpha \bd{c}^\dagger_{\alpha f_1 \sigma_1 2} \bd{c}^{\alpha f_2 \sigma_2 2} =
\sum_\alpha
\bd{d}_{\alpha f_1 \sigma_1} \bd{d}^{\dagger \alpha f_2 \sigma_2} ~~~.
\label{u12}
\end{eqnarray}

The first two equations describe the creation and annihilation of
quark-antiquark pairs. The pairs can be coupled to definite flavor
$(\lambda ,\lambda )$ $=$ $(0,0)$ or $(1,1)$ and spin $S$ $=$ 0 or
1. We shall write, in  this coupling scheme,
$\bd{B}^\dagger_{(\lambda , \lambda )f, S M}$, where $f$ is the
flavor,  $S$ is the spin and $M$ is the spin-projection. The
operators $\bd{B}_{(\lambda , \lambda )f, S M}$ annihilate the
vacuum $|0>$, which is the configuration where the lower state is
completely filled and the upper one is empty.  The operators in
(\ref{u12}) form a U(12) algebra. To simplify the discussion, we
shall restrict to a sub-algebra given by the pair operators
coupled to flavor singlet ($(0,0)$ in the SU(3) notation
\cite{elliott}),

\begin{eqnarray}
\bd{S}_+ & = & \sqrt{6} \bd{B}^\dagger_{(0,0)0, 00} \nonumber \\
\bd{S}_- & = & \sqrt{6} \bd{B}_{(0,0)0, 00} \nonumber \\
\bd{S}_0 & = &  \bd{n}_f - \Omega ~~~,
\label{su2}
\end{eqnarray}
with $n_f=\frac{(n_q + n_{\bar q})}{2}$, where $n_q$ is the number
operator for quarks and $n_{{\bar q}}$ the number operator for the
antiquarks. They form a SU(2) algebra. The model is equivalent to
the Lipkin model \cite{lipkin}, familiar in nuclear physics, with
the difference that the operators are given here by the
combination of quark-antiquark pairs . The addition of an extra
boson level was also discussed in Ref. \cite{schutte} and it is
related to pion effects in nuclei.

As Hamiltonians we shall consider two different types:

\begin{eqnarray}
\bd{H}_{I} & = & 2\omega_f \bd{S}_0 + \omega_b \bd{n}_b + V_1
(\bd{S}_+^2 \bd{b} + \bd{b}^\dagger \bd{S}_{-}^2)
\nonumber \\
\bd{H}_{II} & = & 2\omega_f \bd{S}_0 + \omega_b \bd{n}_b +
V_1 :(\bd{S}_+ + \bd{S}_{-})^2:(\bd{b}^\dagger + \bd{b})
~~~,
\label{ham0}
\end{eqnarray}
with $\bd{S}_0 = \bd{n}_f - \Omega$ and the double
dots indicate {\it normal ordering}.

The Hamiltonian $H_{I}$ exhibits a useful symmetry, i.e. it commutes
with the operator

\begin{eqnarray}
\bd{P} & = & \frac{\bd{n}_f}{2} + \bd{n}_b ~~~.
\label{p}
\end{eqnarray}
This version of the Hamiltonian is similar to the one given in
Ref. \cite{schutte}, except that in \cite{schutte} the SU(2)
operators appear linearly while in (\ref{ham0}) they appear
quadratically. Because a quark-antiquark pair has negative parity,
the operators $\bd{S}_+$ and $\bd{S}_-$  should appear
quadratically to conserve parity.

The states of the model space, belonging to $\bd{H}_{I}$, are
SU(2) states with the additional ordering given by the eigenvalues
of $\bd{P}$. The vacuum state is defined via
$\bd{b}|0>=\bd{S}_{-}|0>=0$. Because the number of fermion pairs
$n_f$ is limited by $2\Omega$ the range of $n_b$ for a fixed value
of $P$ is also limited. Therefore, the matrix representations of
$\bd{H}_{I}$ are finite. A large eigenvalue of $\bd{P}$ implies
that the corresponding configuration has many gluons. In Figure 2
we show the energy of the lowest state for a given  strength  of
the interaction, as a function of the eigenvalue of $\bd{P}$. For
zero interaction, the energy increases monotonically. For large
values of $V_1$ it appears a minimum with a large value of $P$.
This implies that the physical ground state should be a correlated
one, that is to say that the physical ground state will, likely,
be a state with a large number of gluon pairs. Concerning the
dependence upon $V_1$, which is the strength of the interaction
which couples pairs of gluons with pairs of quark-antiquark pairs,
the curves of Figure 2 show minima, which are different from the
perturbative vacuum, for values of  $V_1 \geq 0.035$ GeV. For
larger values of the interaction strength $V_1$, the lowest state
is the one with a large eigenvalue $P$, indicating a gluon
dominated vacuum (for the transitional region, quarks-antiquark
and gluon pairs could appear in the vacuum with comparable
weights). In Figure 3, the energy spectrum of $\bd{H}_{I}$ for
positive parity states is displayed. The parameters used in the
calculations are: $\Omega =9$ (i.e., $n_c =3$ and $n_{fl} =3$),
$\omega_f=\frac{1}{3}$ GeV and $\omega_b=1.6$ GeV. For $V_1=0$ GeV
we obtain the first state at $\frac{4}{3}$ GeV, corresponding to
two quark-antiquark pairs. The next state is the glueball at 1.6
GeV. When the interaction is turned on, the energy changes until
it reaches a "critical" or "transitional" point at $V^c_1=0.035$
GeV. There, a level crossing occurs and the lowest state, for
higher values of $V_1$, has both quark-antiquark and gluon pairs.
Beyond the transitional point, the density of levels increases.
This effect is known from nuclear physics, where the transition
from a spherical nucleus to a deformed one is accompanied by a
significant increase of the density of levels at very low energies
\cite{ring}. For values of the interaction larger than $V^c_1$,
the ground-state expectation-value of the number of
quark-antiquark and gluon pairs increases. For large interaction
strength the number $n_f$ approaches a constant value, reflecting
the Pauli principle, i.e. only a certain number of quarks can
occupy the higher level. This behavior will be discussed in detail
for the case $\bd{H}_{II}$. The relatively high density of states
at low energy, shown in Figure 3, may not be very realistic.
However, the model predicts the appearance of some states at very
low energy. This would correspond to a pion-like structure, which
is also indicative of a collective nature. This problem does not
appear in the case of the Hamiltonian $\bd{H}_{II}$, which does
not commute with $\bd{P}$. The Hamiltonian $\bd{H}_{II}$ has to be
diagonalized in the  whole space, which is infinite dimensional.
The diagonalization can be performed numerically by introducing a
variable cutoff in the number of bosons. We have adopted, as a
criterium for convergence, the stability of the low-energy sector
of the spectrum as a function of the cut-off. The Hamiltonian
$\bd{H}_{II}$ contains all terms which are required by symmetry,
i.e. it includes a term of the form $\bd{S}_+^2\bd{b}$, describing
the annihilation of a gluon pair and the creation of two
quark-antiquark pairs, and also a term $\bd{S}_+^2\bd{b}^\dagger$,
describing ground state correlations. The scattering term
$\bd{S}_+\bd{S}_-(\bd{b}^\dagger + \bd{b})$ appears with a factor
of 2 because fermion lines can be exchanged. Because of the
symmetry in permuting the lines all interactions should have the
same (or at least similar) coupling constant, justifying the use
of only one interaction parameter, $V_1$. Figure 4 shows the
dependence of the energy, of positive (Case (a)) and negative
(Case (b)) parity states, on the strength $V_1$, for the values of
$\omega_f$ and $\omega_b$ given above. The values are referred to
the positive-parity ground-state. Contrary to the case of the
Hamiltonian $\bd{H}_{I}$, the Hamiltonian $\bd{H}_{II}$ does not
show a dense spectrum beyond the phase transition point. In the
transitional region several avoided crossings occur and the
spectrum is richer there, than outside that region. Concerning the
behavior of the low-energy part of the spectrum, it shows, after
the transition point, a negative parity state which is degenerate
with the positive parity ground state. This state, a Goldstone
boson, can be interpreted as a collective, pion-like, state whose
structure can be understood in the framework of coherent states
\cite{hecht}, as it will be discussed later on.

Figure 5 shows the difference of the expectation value of the
number of gluon- and fermion-pairs, as a function of the strength,
$V_1$, of the interaction, for the case of the Hamiltonian
$\bd{H}_{II}$. As it is seen from the figure, the results can be
interpreted in terms of equal population of fermion and gluon
pairs ($V_1 < 0.008$ GeV), fermionic dominance ($0.008 < V_1 <
0.015 $ GeV), and gluonic dominance ($V_1 > 0.015 $ GeV).

The features of the spectrum, and of the ground state occupation
numbers, can be understood in terms of a "phase transition" at
zero temperature, that is to say, in terms of a change in the
correlations induced by the Hamiltonian. A convenient way, to
represent this effect, is to introduce a set of states (coherent
states) and to define an order parameter (the ground state
occupation number associated to a given degree of freedom). The
calculation of the  expectation value of the Hamiltonian, in the
basis of coherent states, and its variation with respect to the
order parameter yields the possible "phases" of the system, as
extremes of the minimization procedure. The set of coherent
states, which we have adopted in our calculations, is defined by
\cite{hecht}

\begin{eqnarray}
| z > & = & | z_f> | z_b> \nonumber \\
| z_f> & = & \frac{1}{(1+|z_f|^2)^{\Omega}}e^{z_f \bd{S}_+}|0>_f \nonumber \\
| z_b> & = & e^{-\frac{|z_b|^2}{2}} e^{z_b \bd{b}^\dagger}|0>_b ~~~,
\label{coh-state}
\end{eqnarray}
where $|0>_f$ and $|0>_b$ are the fermion and the boson vacuum and
$|0>$ is the product vacuum state ($|0>_f \otimes |0>_b$). The
power $\Omega$, of the fermion normalization factor, indicates
that we are working in the SU(2)-spin-representation which
includes, as the lowest state in energy, the fully occupied level
at $-\omega_f$ (see Figure 1).

By defining the complex order parameters $z_f = \rho_f
e^{i\phi_f}$ and $z_b=\rho_b e^{i\phi_b}$, the expectation value
of the Hamiltonian $\bd{H}_{II}$ (\ref{ham0}) is given by

\begin{eqnarray}
<\bd{H}_{II}> & = & -2\Omega\omega_f\frac{(1-\rho_f^2)}{(1+\rho_f^2)} + \omega_b\rho_b^2
\nonumber \\
& + & V_1\left( \frac{4\Omega (2\Omega
-1)\rho_f^2cos(2\phi_f)}{(1+\rho_f^2)^2} +\frac{4\Omega \rho_f^2
(2\Omega +\rho_f^2)}{(1+\rho_f^2)^2} \right) 2\rho_b cos(\phi_b)
~~, \label{potham}
\end{eqnarray}
and it can be regarded as a classical potential in the parametric
space of the amplitudes $\rho$ and phases $\phi$. It shows for
small coupling constants $V_1$ a minimum at $|z_f|=\rho_f=0$,
which represents small departures from a dominant harmonic
(quadratic) potential, and a deformed minimum for a sufficiently
large value of $V_1$ and a given combination of $\phi_f$ and
$\phi_b$. The factor $2\phi_f$, makes the potential invariant
under the change $\phi_f\rightarrow\phi_f+\pi$. Thus, when the
difference in energy of the two minima with respect to the barrier
between them is sufficiently large, there exist two degenerate
states, one with positive and the other with negative parity,
which minimize the expectation value of the Hamiltonian.

The pseudo-scalar particle at zero energy can not be identified
with the pion, because the pion belongs to a flavor $(1,1)$
(octet) representation of the flavor group SU$_f$(3). However, the
fact that the model has a Goldstone boson gives some hope that a
generalized model with open spin and flavor, as indicated at the
beginning, may also exhibit a low lying negative parity state
which can be identified with the pion. This conjecture is
justified because an equivalent interaction, as in Eq.
(\ref{ham0}), for flavor $(1,1)$ spin 0 will exhibit a similar
behavior with respect to a coherent state which includes pairs
with flavor $(1,1)$ spin 0.

In our model the basis for the baryonic states is given by

\begin{eqnarray}
|q^3(q\bar{q})^n(\lambda ,\mu ) f, S M > & \sim & (\bd{S}_+)^n
|q^3, (\lambda \mu ) f, S M> ~~~. \label{barionstates}
\end{eqnarray}
the index $(\lambda ,\mu )f$ refers to the flavor, which is
$(1,1)$ for the octet, etc. The $q^3$ indicates that the state to
the right, on which the operator $\bd{S}_+$ acts, is a pure
three-quark state. The three-quarks state satisfies \cite{bonn}

\begin{eqnarray}
\bd{S}_- |q^3, (\lambda \mu ) f, S M> & = & 0 ~~~. \label{cond1}
\end{eqnarray}

That $\bd{S}_-$ annihilates the three-quark state holds because
the quark-antiquark pair operator contains an antiquark
annihilation operator which anti-commutes with the quark creation
operators of the state on the right and annihilates finally the
correlated vacuum. The basis for the mesons is obtained using
$(\lambda ,\mu ) = (0,0)$ and $S=M=0$. Note that the Hamiltonian
$\bd{H}_{II}$ (and the same is true for $\bd{H}_{I}$) does not
distinguish between different flavors, and, therefore, the flavor-
$(1,1)$ and flavor-$(3,0)$ baryons are degenerate. Part of the
degeneracy can be removed, by introducing terms depending on the
hypercharge and the isospin. In order to remove completely the
degeneracy between the $(1,1)$ and the $(3,0)$ flavor
configurations one has to include, of course, flavor-depending
interactions, as it will be done in a more general formulation of
the present toy model \cite{later}.

The problem for the baryons is completely analogous. Due to the
fact that three quarks minimally occupy the higher level of the
fermion model space, the effective degeneracy, i.e. the number of
configurations available to excite quarks from the lower level, is
$2 \Omega -3$ (since the total number of available states is
$2\Omega$). The factor $3$ is a consequence of the Pauli-blocking.

In Figure 6, the spectrum of the lowest baryonic states, as a
function of the coupling strength $V_1$ and referred to the lowest
positive-parity mesonic state, is shown. After the transition
point, the energy of the baryonic states increases. To obtain
states below 1 GeV, one has to reduce the effective quark mass.
Since we are interested in the trends exhibited by the spectrum,
we shall not fit it to physical masses. The transition point is
slightly shifted to higher values of the coupling constant, due to
the lower value of the degeneracy, otherwise the behavior observed
for the baryon spectrum is similar to the meson case. The
relatively delayed onset of a transition in the baryon spectra
produces a region where the physical vacuum may contain a pair
condensate while the baryonic states may still be built as pure
three quark objects.

In Figure 7 the content of quark-antiquark and gluon pairs in the
first-excited barionic state is shown. The results show a large
contribution from gluon pairs to the baryonic states after the
transition point. This indicate that in a more realistic model and
for sufficiently large interaction strength, some baryon states,
like the proton, may contain a sizeable contribution due to sea
quarks and gluon pairs.

After the analysis of the properties of the low lying spectrum of
the two Hamiltonians (\ref{ham0}), we can proceed to discuss
finite temperature effects, which may be relevant for the
description of the transition from the hadronic phase to the QGP.

\section{The toy model: Finite Temperature Case.}

In this section we shall present the results corresponding to the
finite temperature case. We shall show the main steps related to
the calculation of the partition function, which has been
performed by extending the techniques discussed in
\cite{partition}.

In the limit$V_1=0$, the fermionic sector of the Hamiltonians
${\bd H}_I$ and ${\bd H}_{II}$ reduces to the free Lipkin model,
which consists of two levels, with energies $\pm \omega_f$ and a
degeneracy $(2\Omega)$. Allowed configurations are specified by
all possible arrays of particles and holes, in both levels, and
their degeneracies. Thus, a certain configuration can be specified
by listing the number of occupied (empty) states in the upper
(lower) level. The operators which create (annihilate) these
fermion pairs obey a pseudo-spin block-algebra, for spin
$\frac{1}{2}$. There are in total $2 \Omega$ building blocks. If
$\nu_1$ denotes the number of blocks where both levels are
occupied, $\nu_2$ is the number of blocks where both levels are
empty and $2\tau$ is the number of blocks where either one of the
levels is occupied, the partition function can be written as
\cite{japan}

\begin{eqnarray}
Z (\beta ) & = & \sum_{\tau \nu_1\nu_2}
\frac{(2\Omega )!}{(2\tau )!\nu_1!\nu_2!} \sum_{k=0}^{2\tau} g_k^\tau
I_{\tau -k}
\label{partition1} ~~~,
\end{eqnarray}
where
\begin{eqnarray}
I_J & = & \frac{(2J+1)}{\pi^2} \int d^2z_b \int d^2z_{fJ}
\frac{<z_{fJ}, z_b| e^{-\beta (\bd{H}-\mu \bd{N})}|z_{fJ}, z_b>}{(1+|z_{fJ}|^2)^2} ~~~
\label{partition2}
\end{eqnarray}
with $J=\tau-k$, $\beta = \frac{1}{T}$, and $T$ is the temperature
in units of GeV. The states $|z>=|z>_f \otimes |z>_b$ are the
normalized coherent states \cite{hecht}. The factor $g_k^\tau$ is
the multiplicity of the configuration with pseudo-spin $J=\tau -
k$. For each value of $J$ one should define a coherent state
$|z_{fJ}>$. The coherent state, used in the previous section (the
$T=0$ case), corresponds to the value $J=\Omega$. The chemical
potential $\mu$ multiplies the operator $\bd{N}$, which gives the
total number of particles in the lower and upper level. The
partition function (\ref{partition1}) does not conserve flavor,
color and spin. A multiple projection, to restore these
symmetries, can be carried out, in principle, although it is a
very involved procedure \cite{qgp}. We assume that the volume
occupied by the QGP is large enough so that a projection is not
needed, though we shall restrict to a sub-volume for the
thermodynamic description.

If interactions are added to the Hamiltonian, the partition
function cannot be obtained analytically, in general. If the value
of $\Omega$ is not too large, the partition function
(\ref{partition1}) can be obtained numerically.

We have diagonalized the Hamiltonian $\bd{H}_{I}$ and obtained a
set of eigenvalues for each value of $J$. The parameters of the
Hamiltonian were fixed at $V_1=0$ GeV and $0.04$ GeV, $\Omega =9$,
$\omega_f=\frac{1}{3}$ GeV, and $\omega_b=1.6$ GeV. The partition
function (\ref{partition1}) was calculated  for temperature
$T<0.5$ GeV. Since the value of $\Omega$ is not very large, one
may ask if, at high temperature, more configurations (i.e. larger
values of $\Omega$) should be included. We have checked,
numerically, this effect upon the partition function. Figure 8
shows the value of (\ref{partition1}), for each value of $J \le
\Omega$. As seen from this figure the contributions to the
partition function reach a maximum for a certain value of $J$,
which is smaller than $\Omega$. This result justifies the
approximation with $J>>1$. First and second derivatives of the
partition function (the internal energy and the heat capacity) are
shown in Figures 9 and 10, respectively. The heat capacity, for
$V_1=0$ GeV, shows the Schottky bump \cite{greiner}, typical of a
two level system. The shape of the curve remains unchanged when
the interaction is switched on in ${\bd H}_I$. The increase of the
interaction strength $V_1$ produces a sharper peak in the curve,
indicating a possible second order phase-transition. The results
shown in Figures 9 and 10 have been obtained with the Hamiltonian
${\bd H}_I$. Although we are not showing it in the Figures 9 and
10, the same thermal behavior of the heat capacity is obtained
with the Hamiltonian $\bd{H}_{II}$. However, the values of ($T$
and $V_1$), where the onset of the phase-transition is produced,
are different for the partition functions corresponding to ${\bd
H}_I$ and ${\bd H }_{II}$.

The above discussed results, which have been obtained by
performing a numerical diagonalization are indeed the exact
results of the model. Potentially, they exhibit the desirable
thermodynamical features of QCD. We shall take these results as
reference values for an approximate calculation. The obvious
motivation for such approximate treatment is the generalization to
larger values of the model parameters.

We shall discuss first the treatment of the fermionic sector, i.e.
the integration on the coherent states $\mid z_{\alpha(fJ)}>$. For
large values of $J$, the overlap $<z_\alpha | z_\beta>$ is a
decreasing function of $|\rho_\alpha - \rho_\beta|$. Also in this
limit, it is a strongly oscillating function of the phase
difference $\phi_\alpha-\phi_\beta$. Due to this oscillation we
can write, both for $\bd{H}_{I}$ and $\bd{H}_{II}$

\begin{eqnarray}
<z_{\alpha}|e^{-\beta \bd{H}}|z_{\beta}> & \approx & <z_{\alpha}|
e^{-\beta \bd{H}}|z_{\alpha}><z_{\alpha}|z_{\beta}> ~~~,
\label{approx1}
\end{eqnarray}
where the overlap in the above equation takes into account the off
diagonal behavior for $\alpha \neq \beta$. The expectation value
$<z_\alpha | e^{-\beta {\bd H}} |z_\alpha >$ can be expressed as
an integral product of the form

\begin{eqnarray}
& \frac{(2J+1)}{\pi}\int \frac{d^2z_\alpha}{(1+|z_\alpha |^2)^2)}
<z_\alpha | e^{-\beta \bd{H}} |z_\alpha >  = & \nonumber \\
& \sum_{n=0}^{\infty} \frac{(-\beta)^n}{n!}
\frac{(2J+1)^n}{\pi^n}\int \frac{d^2z_{\gamma_0}...d^2z_{\gamma_{n-1}}}
{\Pi_{k=0}^{n-1}(1+|z_{\gamma_k} |^2)^2}
<z_{\gamma_0} | \bd{H} |z_{\gamma_1} ><z_{\gamma_1} | \bd{H} |z_{\gamma_2} > &
\nonumber \\
& ... <z_{\gamma_{n-1}} | \bd{H} |z_{\gamma_0} > ~~~,
\label{split-H}
\end{eqnarray}
by inserting n-times the unit operator \cite{hecht}. The
Hamiltonian $\bd{H}$ is expressed into two terms
$\bd{H}_0+\bd{H}^\prime$, where $\bd{H}_0$ is the non-interacting
part and $\bd{H}^\prime$ contains the interactions. The k-th term
of the expansion has $k$ matrix elements of $\bd{H}^\prime$ and
$(n-k)$ factors with $\bd{H}_0$. By performing a rearrangement the
contribution of the k-th partition to the integral is written

\begin{eqnarray}
\sum_{n=k}^{\infty} \frac{(-\beta)^n}{n!} \frac{(2J+1)^2}{\pi^2}
& & \int \frac{d^2z_\alpha d^2z_\beta}{(1+|z_\alpha|^2)^2
(1+|z_\beta|^2)^2} \left(
\begin{array}{c}
n \\ k
\end{array}
\right) \nonumber \\
& & <z_\alpha | (\bd{H}^\prime)^n|z_\beta ><z_\beta | (\bd{H}_0)^{n-k}|z_\alpha > ~~~.
\label{div1}
\end{eqnarray}
This result illustrates the binomial character of the k-th
partition. By a re-exponentiation one can write

\begin{eqnarray}
\frac{(2J+1)^2}{\pi^2}\int  \frac{d^2z_\alpha
d^2z_\beta}{(1+|z_\alpha|^2)^2 (1+|z_\beta|^2)^2} <z_\alpha |
e^{-\beta \bd{H}_0}|z_\beta ><z_\beta | e^{-\beta
\bd{H}^\prime}|z_\alpha >~. \label{div2}
\end{eqnarray}
The integral $\int dz_\alpha^2 ~ \int dz_\beta^2$, can be replaced
by a single integral, by using equation (\ref{approx1}) and by
performing the integration on the variable $z_\beta$. The validity
of this approximation is restricted to large values of $J$, whose
validity was shown above. After these approximations
(\ref{split-H}) reads

\begin{eqnarray}
&\frac{(2J+1)^2}{\pi^2} \int \frac{d^2z_\alpha d^2z_\beta}
{(1+|z_\alpha |^2)^2(1+|z_\beta |^2)^2}
<z_\alpha | e^{-\beta \bd{H}_0}|z_\beta ><z_\beta | e^{-\beta \bd{H}^\prime}|z_\alpha >&
\nonumber \\
&\approx  \frac{(2J+1)}{\pi} \int \frac{d^2z_\alpha}{(1+|z_\alpha
|^2)^2} <z_\alpha | e^{-\beta \bd{H}_0}|z_\alpha ><z_\alpha |
e^{-\beta \bd{H}^\prime}|z_\alpha
>~. \label{div3}
\end{eqnarray}

Since $\bd{H}_0=2\omega_f \bd{S}_0$, the matrix element $<z_\alpha
| e^{-\beta \bd{H}_0}|z_\alpha >$ is readily calculated and the
result is

\begin{eqnarray}
<z_\alpha | e^{-\beta 2\omega_f \bd{S}_0} | z_\alpha> & = &
\frac{e^{2\beta\omega_f J}}{(1+|z_\alpha |^2)^{2J}}
(1+|z_\alpha|^2 e^{-2\beta\omega_f})^{2J} ~. \label{fac1}
\end{eqnarray}

As a check on the consistency of these approximations, it is
verified that by setting ${\bd H}^{\prime}=0$ and from the above
equation, the integration of (\ref{partition2}) gives a result
which is identical to the one of \cite{japan}.

Concerning the matrix element of the interaction ${\bd
H}^{\prime}$, we have adopted the following approximation

\begin{eqnarray} <z_\alpha | e^{-\beta \bd{H}^\prime} | z_\alpha >
& \approx & e^{-\beta <z_\alpha | \bd{H}^\prime | z_\alpha >} ~,
\label{approx2}
\end{eqnarray}
which is valid when the temperature is high and/or the interaction
coupling constant is small. It corresponds to the factorization
$<z_\alpha | (\bd{H}^\prime)^n| z_\alpha >$ $\approx$ $(<z_\alpha
| \bd{H}^\prime| z_\alpha >)^n$, a result which can be reproduced,
by assuming (\ref{approx1}) and inserting $n$-th unit operators
between factors $\bd{H}^\prime$. As said before, the applicability
of the procedure is limited to relatively small values of the
interaction strength $V_1$ and relatively large values of the
temperature.

Let us now turn the attention to the bosonic degrees of freedom.
The exponent of Eq. (\ref{approx2}) is a linear combination of the
boson operators $\bd{b}^\dagger$ and $\bd{b}$. The coefficients
are given by the expectation values of $\bd{S}_{\pm}^2$ or
$\bd{S}_+\bd{S}_-$ (see appendix A), which are functions of the
complex variables $z_\alpha$ and $z^*_\alpha$. The normalized
boson coherent state is given by \cite{hecht}

\begin{eqnarray}
|z_b> & = & e^{-\frac{|z_b|^2}{2}} e^{z_b \bd{b}^\dagger} | 0 > ~.
\label{boson-coh}
\end{eqnarray}
The volume element of the complex integral is $\frac{d^2z_b}{\pi}$
(in Ref. \cite{hecht} the coherent states are not normalized and
therefore the volume element has an extra factor $e^{-|z_b|^2}$).
The calculation involves an integral of the type

\begin{eqnarray}
\frac{1}{\pi} \int d^2z_b <z_b| e^{-\beta\omega_b \bd{n}_b
+a_1\bd{b}^\dagger + a_2 \bd{b}} | z_b> ~. \label{int-boson}
\end{eqnarray}
The coefficients $a_k$, which depend on the expectation value of
powers of the SU(2) generators and which are proportional to the
interaction strength $V_1$, have a common value for the model
Hamiltonian $\bd{H}_{II}$ but they differ for the case of
$\bd{H}_{I}$.

In order to evaluate the expectation value which appears in
(\ref{int-boson}), the Baker-Campbell-Hausdorff formula is applied
\cite{hecht,mertzbacher}. Notice that one cannot apply the
approximations described for the fermion sector, because the
overlaps $<z_b | z^\prime_b>$ show a broader dependence on
$\rho_b-\rho_b^{\prime}$ and $\phi_b-\phi_b^{\prime}$. The result
is

\begin{eqnarray}
e^{-\beta\omega_b \bd{n}_b +a_1\bd{b}^\dagger + a_2 \bd{b}} & = &
e^{\xi_1 \bd{b}^\dagger} e^{\xi_2 \bd{n}_b} e^{\xi_3 \bd{b}}
e^{-K}~, \label{bch}
\end{eqnarray}
with

\begin{eqnarray}
K & = \xi_1\xi_3\left[ e^{-\xi_2}\left( \frac{1}{\xi_2^2}+\frac{1}{\xi_2}
\right) -\frac{1}{\xi_2^2}\right] ~, \nonumber \\
\xi_1(I) & = \frac{2\beta V_a}{2+\beta\omega_f} ~,~
\xi_1(II) & = -\frac{2\beta V_c}{2+\beta\omega_b} \nonumber \\
\xi_2(I) & = -\beta\omega_b ~,~
\xi_2(II) & = -\beta\omega_b \nonumber \\
\xi_3(I) & = -\frac{\beta^2\omega_b V_b}{(1-e^{-\beta\omega_b})}
e^{-\beta\omega_b} ~,~ \xi_3(II) & = -\frac{\beta^2 V_c
\omega_b}{(1-e^{-\beta\omega_b})} e^{-\beta\omega_b}~, \label{xi}
\end{eqnarray}
and where

\begin{eqnarray}
V_a & = &  V_1 < z_\alpha |\bd{S}_-^2| z_\alpha > \nonumber \\
V_b & = &  V_1 < z_\alpha |\bd{S}_+^2| z_\alpha > \nonumber \\
V_c & = &  V_1 < z_\alpha |:(\bd{S}_+ + \bd{S}_- )^2:| z_\alpha >
~. \label{vavb}
\end{eqnarray}
The indices $I$ and $II$ refer to the two different model Hamiltonians.

Because the exponential function which contains the operator
$\bd{b}$ is acting on the coherent state $|z_b>$, the operation is
well defined and it gives a factor $e^{\xi_3 z_b}$. The same holds
for the exponential function which contains the operator
$\bd{b}^\dagger$ and it gives $e^{\xi_1 z^*_b}$. The expectation
value of the operator which contains $\bd{n}_b$ reads

\begin{eqnarray}
<z_b| e^{\xi_2 \bd{n}_b} | z_b> & = & e^{-|z_b|^2}
e^{|z_b|^2e^{\xi_2}} ~. \label{nb-exp}
\end{eqnarray}
Finally, the integration over the complex variable $z_b$ yields

\begin{eqnarray}
\frac{1}{\pi} \int d^2z <z_b | e^{\xi_1 \bd{b}^\dagger} e^{\xi_2
\bd{n}_b} e^{\xi_3 \bd{b}} | z_b> & = &
\frac{e^{\frac{\xi_1\xi_3}{1-e^{\xi_2}}}}{1-e^{\xi_2}} ~.
\label{int-comp}
\end{eqnarray}
In the following we shall construct the final expression of the
partition function, after the above introduced approximations. We
shall restrict to the case of $\bd{H}_{I}$, in order to compare
the results of the approximations with the numerical (exact)
results. A similar analysis can be performed for the case of
$\bd{H}_{II}$ \cite{later}. The remaining integration on the
complex variable $z_\alpha$, of the fermion part, leads to

\begin{eqnarray}
\left( \frac{1}{1-e^{-\beta \omega_b}} \right) (2J+1)
\int_{0}^{\infty} dx \frac{e^{2\beta \omega_f J}}{(1+x)^{2J+2}}
(1+x e^{-2\beta\omega_f})^{2J} e^{\beta 16F(\beta) V_1^2
J^4\frac{x^2}{(1+x)^4}} ~, \label{integral}
\end{eqnarray}
with

\begin{eqnarray}
F(\beta) = 2 \left(\frac{1}{\omega_b}\right)^2 \frac{\beta
\omega_b \left((\beta\omega_b)^2 - e^{-\beta\omega_b}-
e^{\beta\omega_b} + \beta\omega_b e^{\beta\omega_b}
-\beta\omega_b+2 \right)}
{(2+\beta\omega_b)(e^{\beta\omega_b}-1)(1-e^{-\beta\omega_b})},
\nonumber
\end{eqnarray}
after substituting $x=\rho^2$. $F(\beta)$ is a smooth function of
$\beta$, which approaches the limit $\frac{2\beta}{\omega_b}$ for
large $\beta$. The integral, (\ref{integral}), is a function of
the pseudo-spin $J$ and its argument is the product of exponential
functions of positive and negative functions of J. That the
integral can be represented by the integrand at $J=J_0$, where
$J_0$ is the value that maximizes it, can be easily seen by
setting $V_1=0$.

Let us call $g(J)$ the ratio between the integral (\ref{integral})
with $V_1 \neq 0$ and the same integral with $V_1=0$, and $J_0$
the value of $J$ which maximizes the integral. We can write the
Taylor expansion of $g(J)$ as

\begin{eqnarray}
g(J) & = & \sum_n \frac{1}{n!} \left( \frac{\partial^n
g}{\partial{J^n}}\right)_{J=J_0} (J-J_0)^n ~. \label{taylor}
\end{eqnarray}
By using the identity

\begin{eqnarray}
J^nX^{2J} & = & \frac{1}{2^n}\left( \frac{\partial^n}{\partial
X^n} X^{2J}\right) X^n ~, \label{formula-j}
\end{eqnarray}
with $X=e^{\beta\omega_f}$, the expansion in terms of $(J-J_0)$ of
$g(J)$ transforms into an expansion in the derivatives
$\frac{\partial}{\partial X}$. Thus, the sum (\ref{taylor}) is of
the form $\sum_n D^n(J_0,X)Z_0$, where $Z_0$ is the partition
function for the non-interacting case \cite{japan} and $D$ is the
differential operator $(\frac{X}{2}\frac{\partial}{\partial X}-
J_0)$. We choose $J_0$ such that the $ (\frac {X}{2} \frac
{\partial} {\partial X} - J_0)Z_0 =0$. The value of $J_0$,
obtained in this way, is given by

\begin{eqnarray}
J_0 & \approx & \Omega \left(\frac{X^2Y}{(X+Y)(1+XY)}
\right)(1-\frac{1}{X^2}) ~, \nonumber
\end{eqnarray}
where $Y$ is the fugacity \cite{japan}. This expression approaches
$J_0=\Omega$ for $\beta\rightarrow\infty$ and it gives small
values of $J_0$ for $\beta\rightarrow 0$. After these
approximations the partition function can be written as

\begin{eqnarray}
Z & = & Z_f Z_b Z_{int} ~, \label{part-final}
\end{eqnarray}
where $Z_f$ and $Z_b$ are the partition functions for free
fermions and free bosons, respectively, and $Z_{int}$ is the
contribution due to the interactions. For $\beta \rightarrow 0$
$Z_{int} \rightarrow 1$ by construction, and the high temperature
limit is automatically obtained.

In what follows we choose the same values of interaction strength
as done above when the partition function was calculated numerically.

In Figures 9 and 10 we show the internal energy and the heat
capacity respectively, as a function of the temperature, obtained
with the approximate partition function (\ref{part-final}). As it
can be seen from the curves, the approximation works reasonable
well for high temperatures, $T > 0.15$ GeV. At low temperatures
the curves, representing the approximated values, deviate
significantly from the exact results.

The behavior of the curves of Figure 9, illustrates the agreement
between the approximated and the exacts results. It is valid only
for the high temperature region. At low $T$, the main contribution
to the partition function comes from the ground state. The lowest
energy is the one of the configuration where all states in the
lower level are occupied while the ones in the upper level are
empty. This corresponds to the case $(2\tau)=2\Omega$,
$\nu_1=\nu_2=0$. The partition function reduces to one term, given
by the integral (\ref{integral}) with $J=\Omega$, and the energy
goes to $-\frac{2V_1^2\Omega^4}{\omega_b}$ as $T$ goes to zero.
This value, is of the order of $-13$ GeV if one uses the already
given parameters ($V_1=0.04$ GeV, $\Omega=9$ and $\omega_b=1.6$
GeV). The exact calculation yields $-2$ GeV. This deviation is
caused by the approximation (\ref{approx2}).

Note that, in a more realistic context, the transition to the QGP
is believed to take place around $T=0.165$ GeV or $T=0.270$ GeV
\cite{qmd}. For these temperatures our approximate results are
acceptable. Also, for these temperatures there is still a sizeable
difference between the results obtained with interactions and
without interactions. It shows also that the interactions cannot
be neglected in the high temperature regime.

Finally, we like to discuss the dependence of $T$ on the chemical
potential $\mu$, when the pressure of the system is equal to the
bag pressure $B=0.145$GeV \cite{greiner}. To find such a
dependence one has to introduce a volume. We consider an
elementary volume given by the size of a hadron ($\approx
1$fm$^3$), as done in \cite{qgp}. The pressure is then given by
the ratio of the internal energy and this volume. The result is
depicted in Figure 11. Without interactions ($V_1=0$) we reproduce
the results of Ref. \cite{greiner}. By turning on the interaction
($V_1= 0.04 $ GeV), the chemical potential increases. This effect
shows that the correlated vacuum state is dominated by gluon pairs
and it also has contributions from quark-antiquark pairs.

\section{Conclusions}

In this work we have introduced the essentials of a toy model for
QCD. The model consists of two levels with energies $\pm
\omega_f$, which describe the fermion degrees of freedom, and
gluons are introduced via a level of positive energy which can be
filled by gluon pairs with spin zero. The gluon pairs are treated
as bosons. Two different Hamiltonians are discussed. The first
one, ($\bd{H}_I$), commutes with the symmetry operator $\bd{P}$, a
property which allows us to calculate the energy spectrum easily.
However, this Hamiltonian contains only a certain type of
interaction terms which do not include ground state correlations.
The second Hamiltonian ($\bd{H}_{II}$) does contain terms which
produce ground state correlations. Due to the symmetry of the
vertices, respect to the exchange of fermion and boson lines, all
terms entering in $\bd{H}_{II}$ have the same interaction
strength. We have shown that the corresponding spectrum exhibits a
phase transition, depending on the interaction. For small values
of the interaction $V_1$, fermion pairs and gluon pairs equally
populate the ground state, for intermediate values of $V_1$ the
physical vacuum is described by quark-antiquark pairs while for
larger values of $V_1$ gluon pairs dominate. In the gluon
dominated phase the spectrum has a degeneracy of the ground state,
given by one positive- and one negative-parity state. This
property gives some hope that a more general version of the model,
with open flavor and spin channels, may show a Goldstone boson (
in the flavor octet $(1,1)$ SU(3) notation), if the strength of
the corresponding interaction is large enough. The appearance of
the gluon and quark-antiquark condensate and of the Goldstone
boson, may be easily described by using coherent states, as we
have shown for the present version of the model. The same approach
may be useful for the case of  more general models of
non-perturbative QCD.

Baryons were also considered  in the model (see Eq.(8)) and due to
Pauli-blocking effect the effective degeneracy $\Omega$ decreases.
Also, the transition point to a condensate of pairs of
quark-antiquarks and gluons is shifted to larger values of the
interaction strength. There is a regime where mesons are already
in a condensate phase while the baryons can still be treated as
three quark systems. We have investigated finite temperature
effects, by constructing the partition function of the model, both
exactly and approximately. We have shown that the use of coherent
states makes it possible to introduce approximations in a
controlled way. The results, for the internal energy, heat
capacity and the equation of state of the system at the bag
pressure, are in agreement with previous calculations. To
summarize, the model is able to describe characteristic features
of QCD at low and at high temperature. This gives some hope that,
in  a more general version of it with open flavor and color, it
may describe the hadron spectrum at low energy and the transition
to the QGP, as well.

\section*{Appendix A}

We want to calculate the expectation values of the operators
$\bd{S}_0, (\bd{S}_+^2 +\bd{S}_-^2) $
and $\bd{S}_+ \bd{S}_-$ using coherent states. We adopted, as a suitable representation,
the normalized coherent states  \cite{hecht}

\begin{eqnarray}
\mid z_f> = \frac{1}{(1+|z_f|)^J}e^{z_f \bd{S}_+}\mid J,-J>
\nonumber
\end{eqnarray}
where $\mid J,-J>$ is the eigenstate of the pseudo-spin algebra
with the absolute value of the spin $J$ and its projection $M=-J$.

For the operator $\bd{S}_0$ we have

\begin{eqnarray}
<z_f\mid \bd{S}_0 \mid z_f> =
\frac{1}{(1+|z_f|^2)^{2J}} <J,-J\mid (e^{z_f^{\star} \bd{S}_-}
\bd{S}_0 e^{z_f \bd{S}_+})\mid J,-J>
\nonumber
\end{eqnarray}

Now, since
$[\bd{S}_-,\bd{S}_0]=\bd{S}_-$ and

\begin{eqnarray}
e^{\bd{A}} \bd{O} e^{-\bd{A}} & = & \bd{O} +\frac{1}{1!}[\bd{A},\bd{O}] +
\frac{1}{2!}[\bd{A},[\bd{A},\bd{O}]] +... ~~~,
\end{eqnarray}
we arrive at

\begin{eqnarray}
e^{z_f^{\star} \bd{S}_-} \bd{S}_0 e^{z_f\bd{S}_+} =
(\bd{S}_0+z_f^{\star} \bd{S}_-)e^{z_f^{\star}\bd{S}_-}e^{z_f \bd{S}_+}
\nonumber
\end{eqnarray}

This yields

\begin{eqnarray}
<z_f\mid \bd{S}_0 \mid z_f> =
\frac{1}{(1+|z_f|^2)^{2J}} <J,-J\mid
(\bd{S}_0+z_f^{\star} \bd{S}_-)e^{z_f^{\star} \bd{S}_-}
e^{z_f \bd{S}_+}\mid J,-J> =
\nonumber
\end{eqnarray}

\begin{eqnarray}
\frac{1}{(1+|z_f|^2)^{2J}}(-J+z_f^{\star} \frac{\partial}
{\partial z_f})
<J,-J\mid e^{z_f^{\star} \bd{S}_-}e^{z_f \bd{S}_+}\mid J,-J>
\nonumber
\end{eqnarray}

To calculate the normalization,
$<J,-J\mid e^{z_f^{\star} \bd{S}_-}e^{z_f \bd{S}_+}\mid J,-J>$,
we write \cite{hecht}

\begin{eqnarray}
(1+|z_f|)^J \mid z_f> = e^{z_f \bd{S}_+}\mid J,-J>
=  \sum_{n=0}^{2J}\frac{|z_f|^n}{\sqrt{n!}}
\sqrt{\frac{(2J)!}{(2J-n)!}}\mid J,n>
\nonumber
\end{eqnarray}

leading to
$<J,-J\mid e^{z_f^{\star} \bd{S}_-}e^{z_f \bd{S}_+}\mid J,-J> =
(1+|z_f|^2)^{2J}$.

With this result, the expectation value of $\bd{S}_0$ reads
\begin{eqnarray}
<z_f\mid \bd{S}_0 \mid z_f> & = &
\frac{1}{(1+|z_f|^2)^{2J}}(-J+z_f^{\star} \frac{\partial}
{\partial z^\star_f}) (1+|z_f|^2)^{2J} \nonumber \\
& = & -\frac{J (1-|z_f|^2)}{1+|z_f|^2}
\nonumber
\end{eqnarray}

Next we are going to calculate the expectation value of the operator
$\bd{S}_+^2$.

\begin{eqnarray}
<z_f \mid \bd{S}_+^2\mid z_f> =
\frac{1}{(1+|z_f|^2)^{2J}}
<J,-J \mid (e^{z_f^{\star} \bd{S}_-}\bd{S}_+^2 e^{z_f \bd{S}_+})
\mid J,-J> =
\nonumber
\end{eqnarray}

\begin{eqnarray}
\frac{1}{(1+|z_f|^2)^{2J}} \frac{\partial^2}{\partial z_f^2}
<J,-J\mid (e^{z_f^{\star} \bd{S}_-} e^{z_f\bd{S}_+})
\mid J,-J> =
\nonumber
\end{eqnarray}

\begin{eqnarray}
\frac{1}{(1+|z_f|^2)^{2J}} \frac{\partial^2}{\partial z_f^2}
(1+|z_f|^2)^{2J} =
\frac{2J (2J-1) (z_f^{\star})^2}{(1+|z_f|^2)^{2}}.
\nonumber
\end{eqnarray}

Therefore

\begin{eqnarray}
<z_f \mid (\bd{S}_+^2 + \bd{S}_-^2)\mid z_f> =
\frac{2J(2J-1)((z_f^{\star})^2+z_f^2)}{(1+|z_f|^2)^2}
\nonumber
\end{eqnarray}

To calculate the expectation value of $\bd{S}_+ \bd{S}_-$
we use the identity
$\bd{S}_+ \bd{S}_- = \bd{S}^2 - \bd{S}_0^2 + \bd{S}_0$, and from the above obtained results
we find
\begin{eqnarray}
<z_f \mid \bd{S}_0^2 \mid z_f> = \frac{1}{(1+|z_f|^2)^{2J}}
(-J+z_f^{\star}\frac{\partial}{\partial z_f^{\star}})^2 (1+|z_f|^2)^{2J} =
\nonumber
\end{eqnarray}

\begin{eqnarray}
J^2 - \frac{4J^2 |z_f|^2}{1+|z_f|^2} +
\frac{2J |z_f|^2}{1+|z_f|^2} +
\frac{2J(2J-1)|z_f|^4}{(1+|z_f|^2)^2} =
J^2 - \frac{2J(2J-1)|z_f|^2}{(1+|z_f|^2)^2},
\nonumber
\end{eqnarray}

and

\begin{eqnarray}
<z_f \mid \bd{S}_+ \bd{S}_- \mid z_f> =
<z_f \mid (\bd{S}^2 -\bd{S}_0^2 + \bd{S}_0)\mid z_f> =
\nonumber
\end{eqnarray}

\begin{eqnarray}
J(J+1) -(J^2 - \frac{2J(2J-1)|z_f|^2}{(1+|z_f|^2)^2})
-\frac{J(1-|z_f|^2)}{1+|z_f|^2} =
\frac{2J |z_f|^2(2J+|z_f|^2)}{(1+|z_f|^2)^2}
\nonumber
\end{eqnarray}

Finally, in the limit of large $J$ expressions like ($2J-k$) are
approximated by $2J$.

\section*{Acknowledgments}
We acknowledge financial support through the CONACyT-CONICET
agreement under the project name {\it Algebraic Methods in Nuclear
and Subnuclear Physics} and from CONACyT project number 32729-E.
(S.J.) acknowledges financial support from the {\it Deutscher
Akademischer Austauschdienst} (DAAD) and SRE, (S.L) acknowledges
financial support from DGEP-UNAM.

\newpage
\begin{center}
{\bf FIGURE CAPTIONS:} \\
\end{center}

\begin{picture}(100,200)(9,9)
\psfig{file=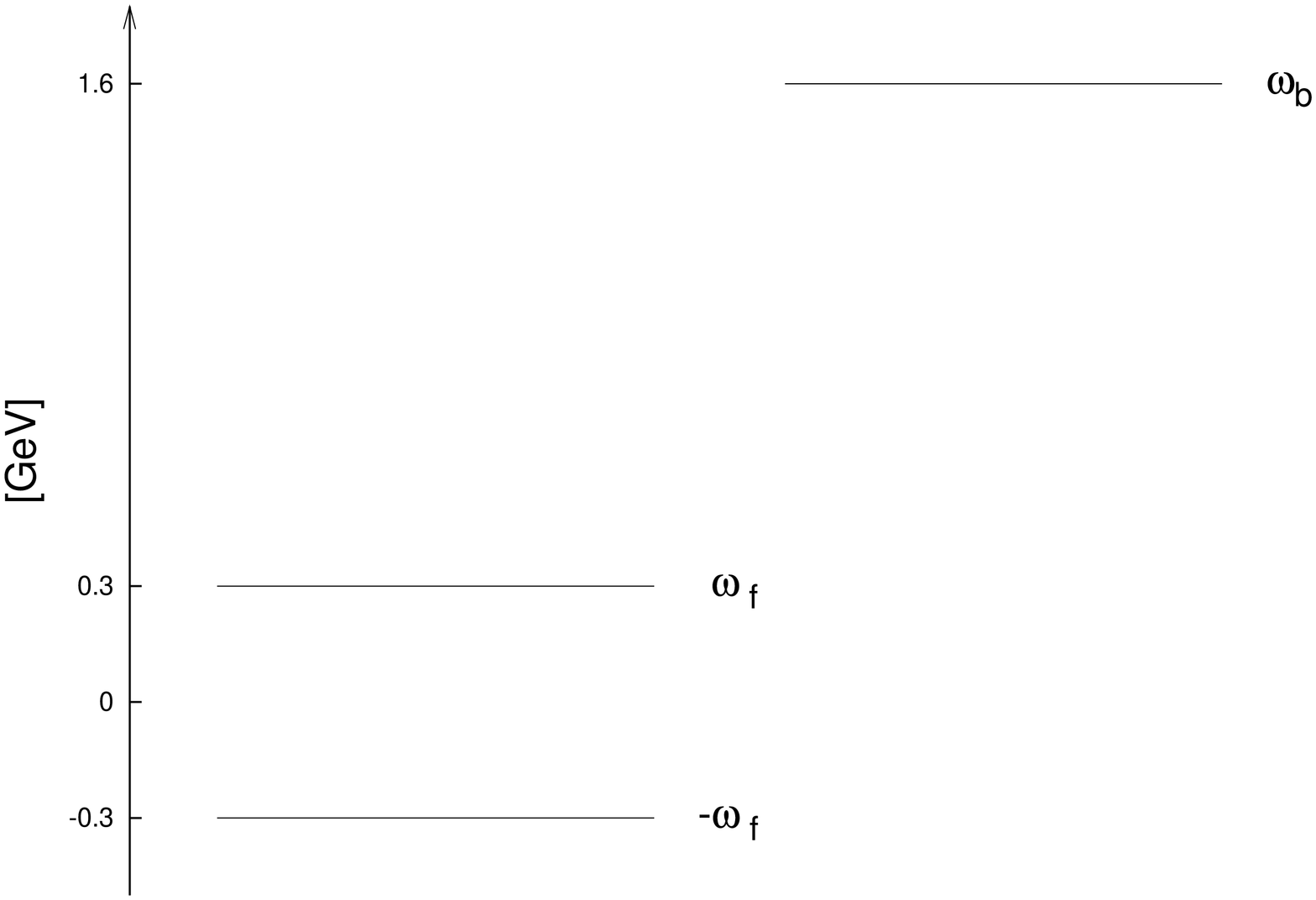,width=10.0cm,height=7cm,angle=270}
\end{picture}
\vskip 1cm \noindent {\bf Figure 1}: Schematic representation of
the model space. The fermion levels are indicated by their
energies $\pm \omega_f$. The gluon-pairs are represented by the
level at the energy $\omega_b$.\\

\newpage
\begin{picture}(100,200)(9,9)
\psfig{file=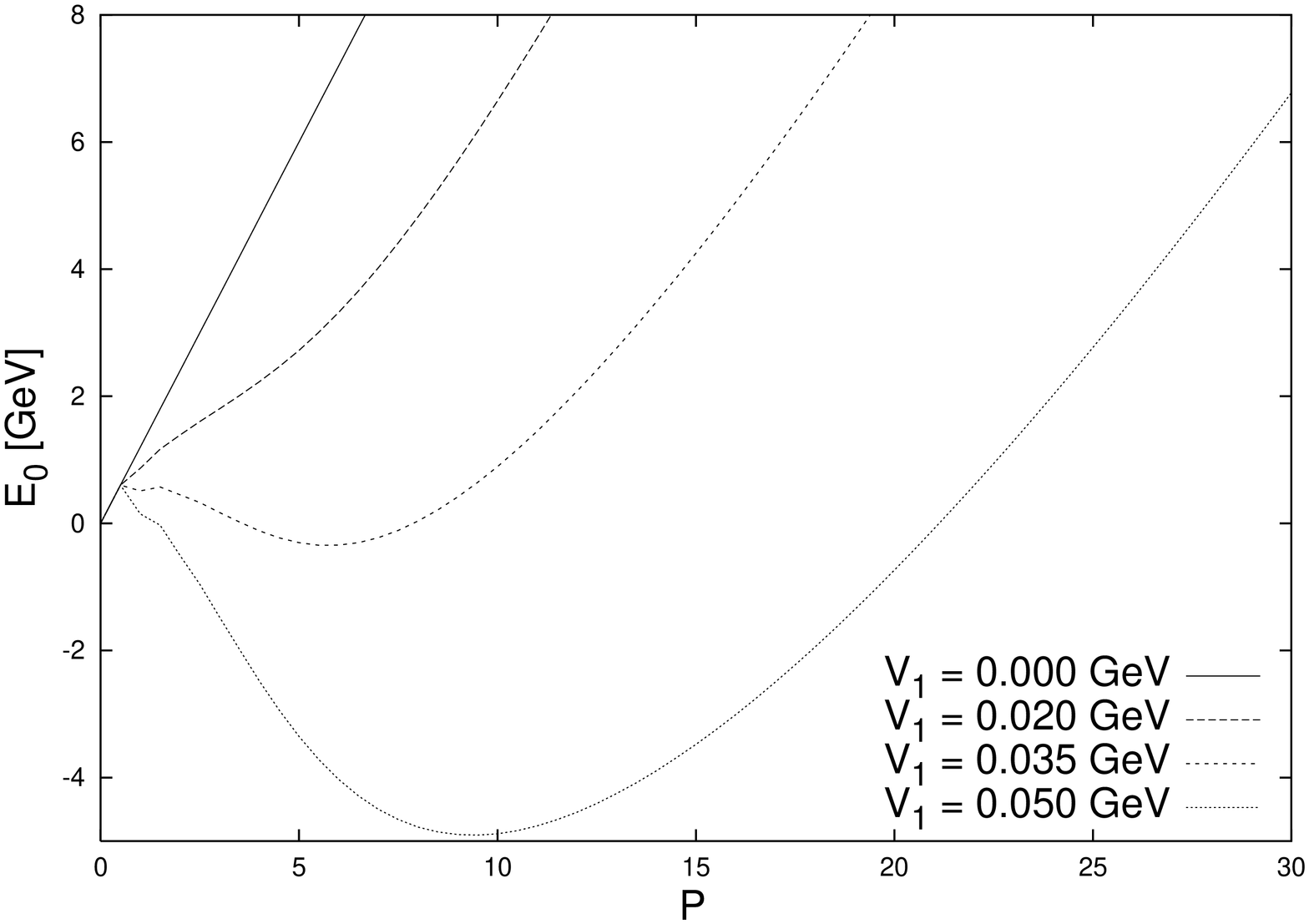,width=7.0cm,height=7cm,angle=-90}
\end{picture}

\vskip 1cm \noindent {\bf Figure 2}:  The energy of the ground
state, $E_0$, in units of GeV, as a function of $P$, for several
values of the parameter $V_1$. The values correspond to the
calculations  performed with the Hamiltonian $\bd{H}_{I}$. Note
the occurrence of a nontrivial minima when $V_1 \geq 0.035$ GeV.

\newpage
\begin{picture}(100,200)(9,9)
\psfig{file=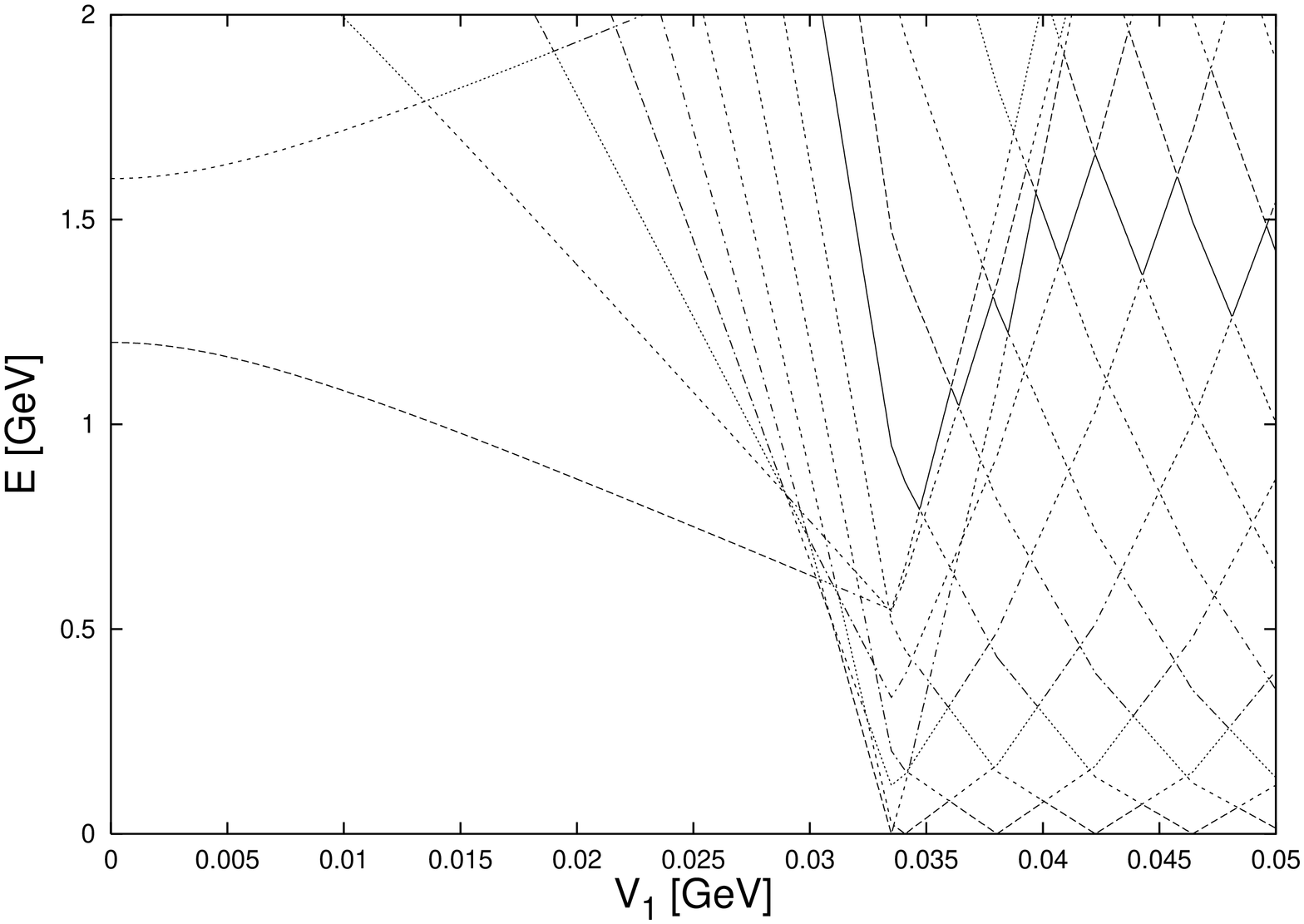,width=7.0cm,height=7cm,angle=-90}
\end{picture}

\vskip 1cm \noindent {\bf Figure 3}: The spectrum of the model
Hamiltonian $\bd{H}_{I}$, for positive parity states, as a
function of the coupling parameter $V_1$. Note the crossing of
excited states with the perturbative ground state at about $V_1
\approx 0.035$ GeV.

\newpage
\begin{picture}(100,200)(9,9)
\psfig{file=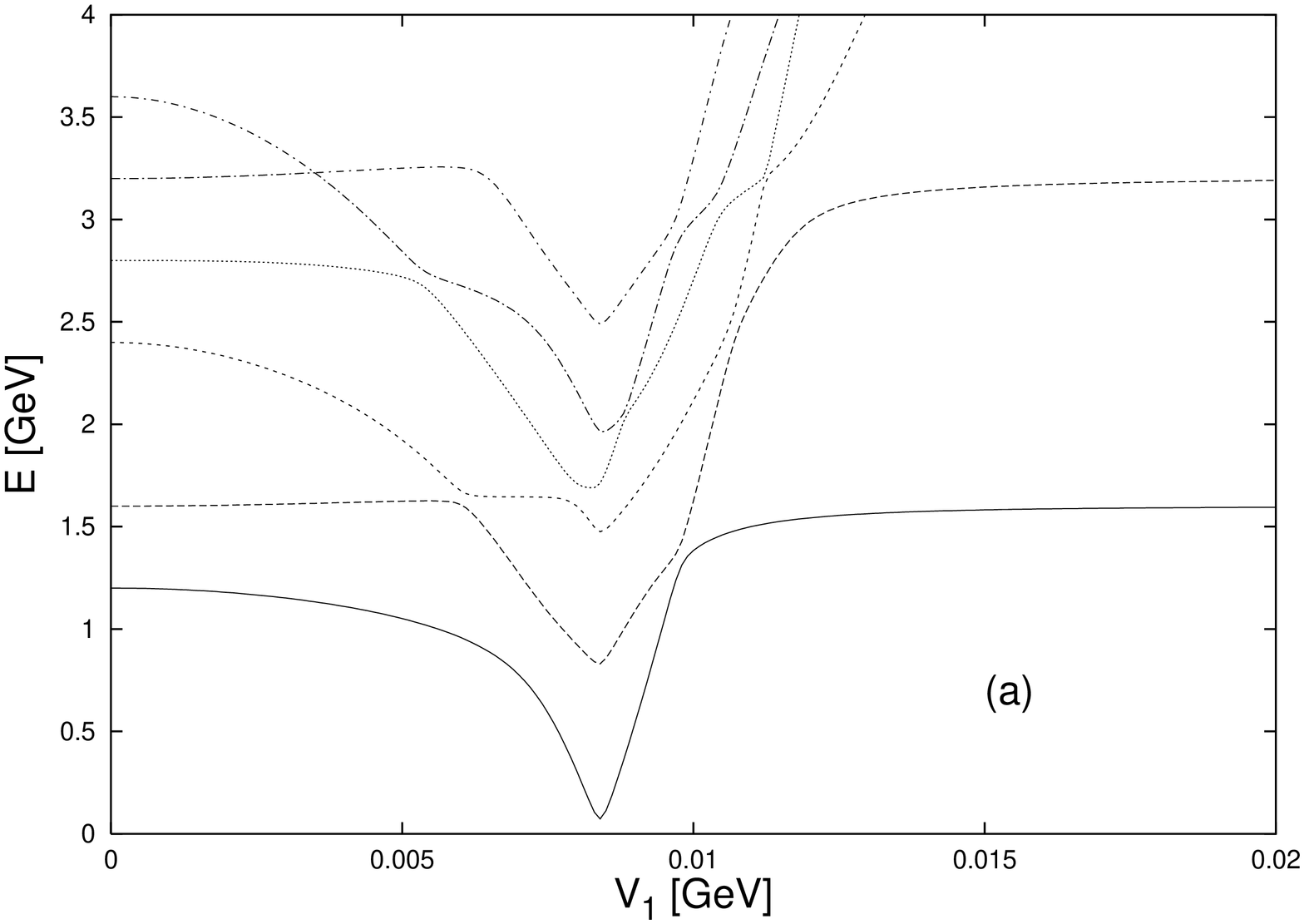,width=10.0cm,height=7cm,angle=270}
\end{picture}

\begin{picture}(100,200)(9,30)
\psfig{file=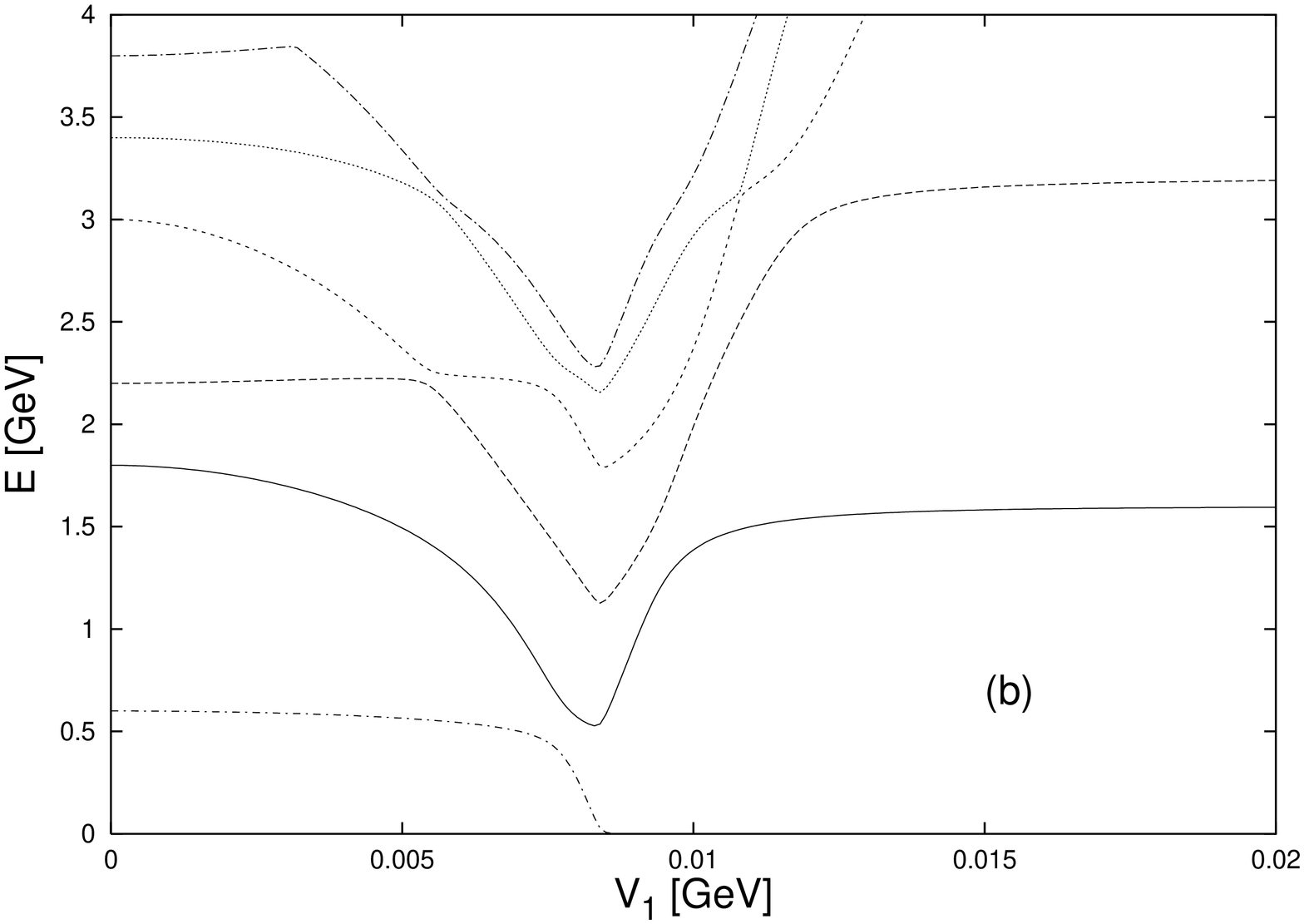,width=10.0cm,height=7cm,angle=270}
\end{picture}
\vskip 1.5cm \noindent {\bf Figure 4}: The energy spectrum of
Hamiltonian $\bd{H}_{II}$ for positive (Case (a)) and negative
parity (Case (b)) states, as a function of the coupling strength
$V_1$. The on set of the phase transition takes place for lower
values of $V_1$, as compared with the results corresponding to the
Hamiltonian $\bd{H}_{I}$ (see Figure 3).

\newpage
\begin{picture}(100,200)(9,9)
\psfig{file=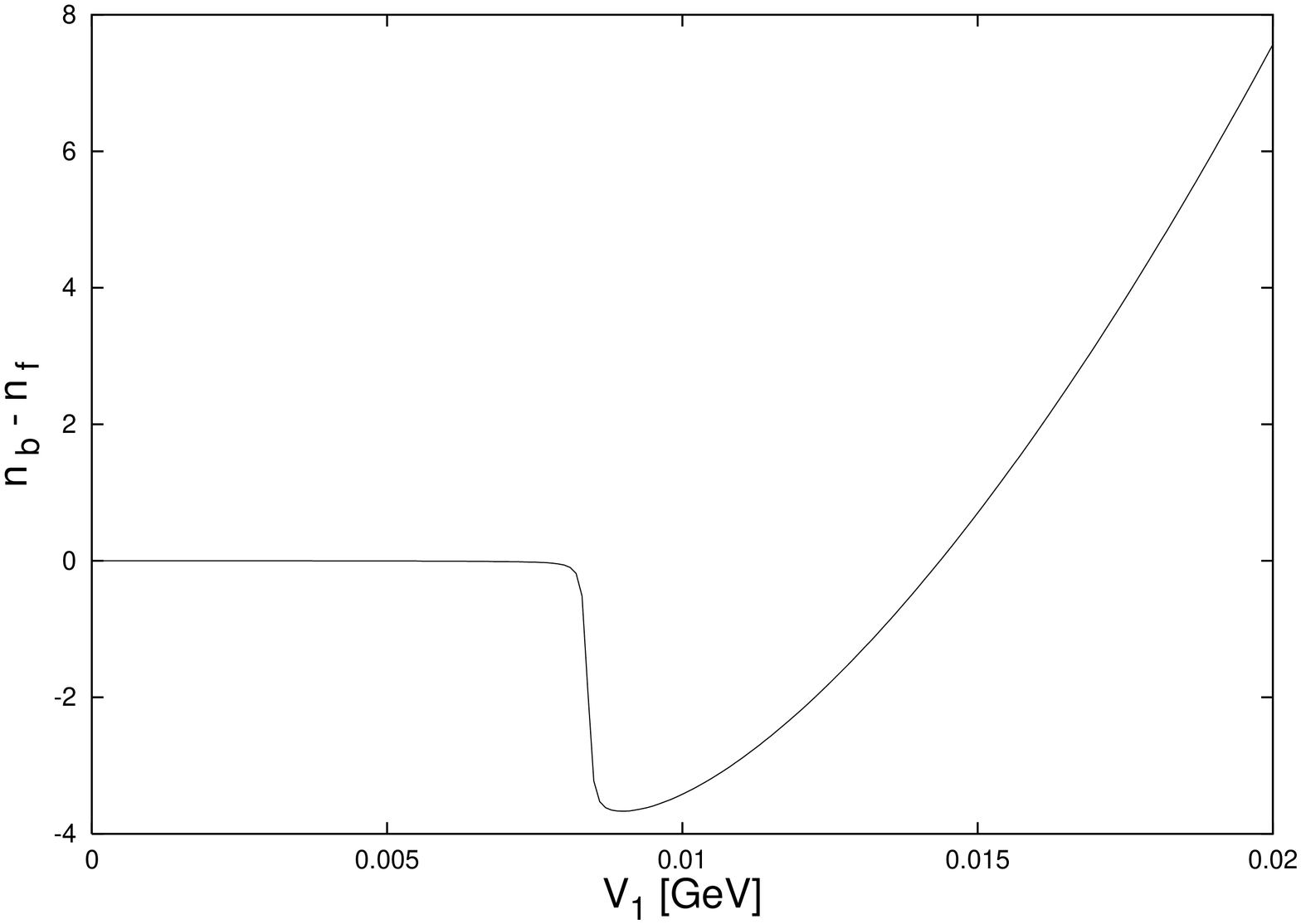,width=10.0cm,height=7cm,angle=270}
\end{picture}

\vskip 1cm \noindent {\bf Figure 5}: The difference between the
vacuum expectation values of the number of gluon pairs $n_b$ and
fermion pairs $n_f$ is shown, as a function of $V_1$, and for the
case of the Hamiltonian $\bd{H}_{II}$.

\newpage
\begin{picture}(100,200)(9,9)
\psfig{file=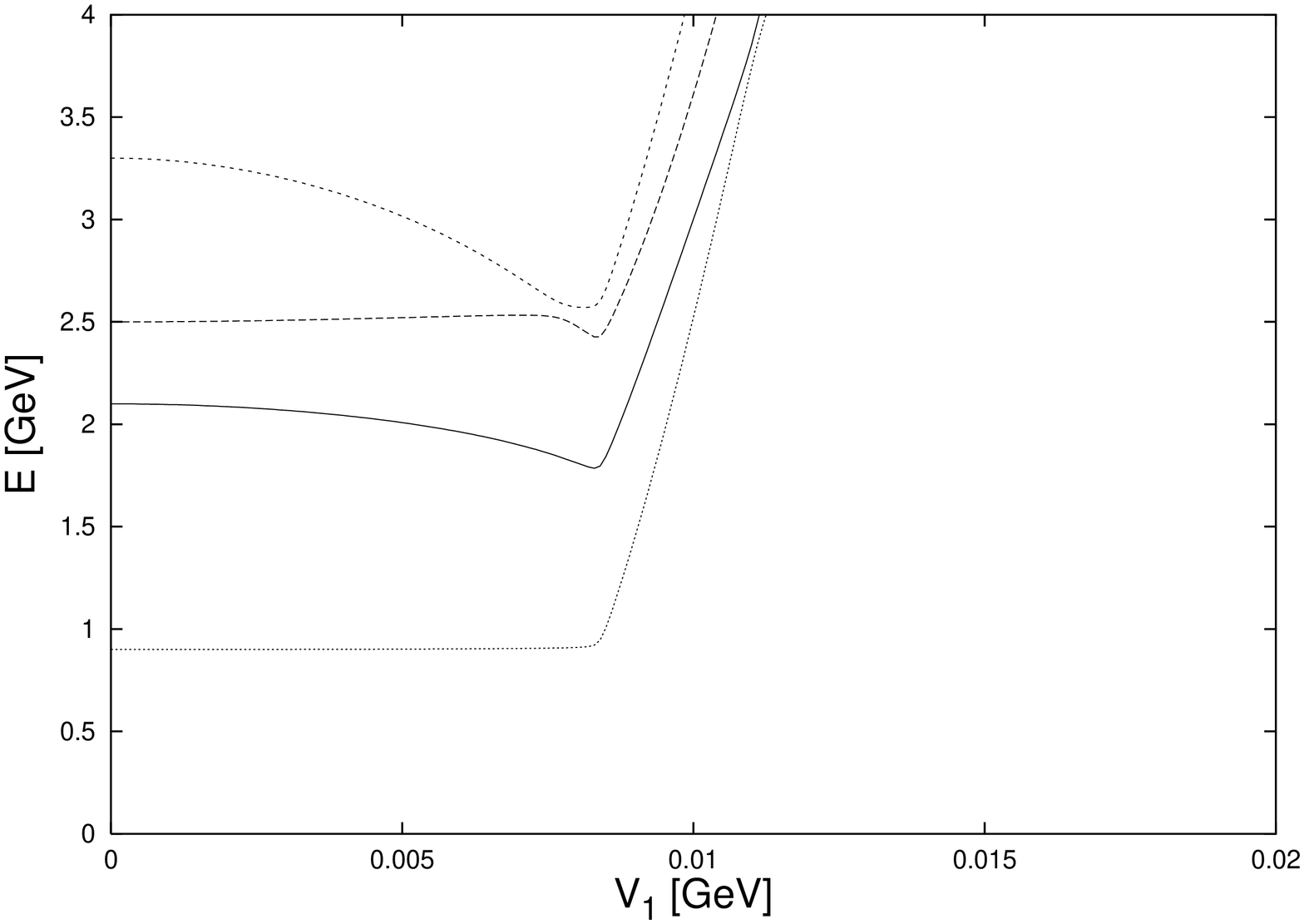,width=10.0cm,height=7cm,angle=270}
\end{picture}

\vskip 1cm \noindent {\bf Figure 6}: The spectrum of the lowest
baryonic states as a function of the interaction strength. The
values of the excitation energies are taken with respect to the
ground state of the mesonic sector. The Hamiltonian $\bd{H}_{II}$
was used.

\newpage
\begin{picture}(100,200)(9,9)
\psfig{file=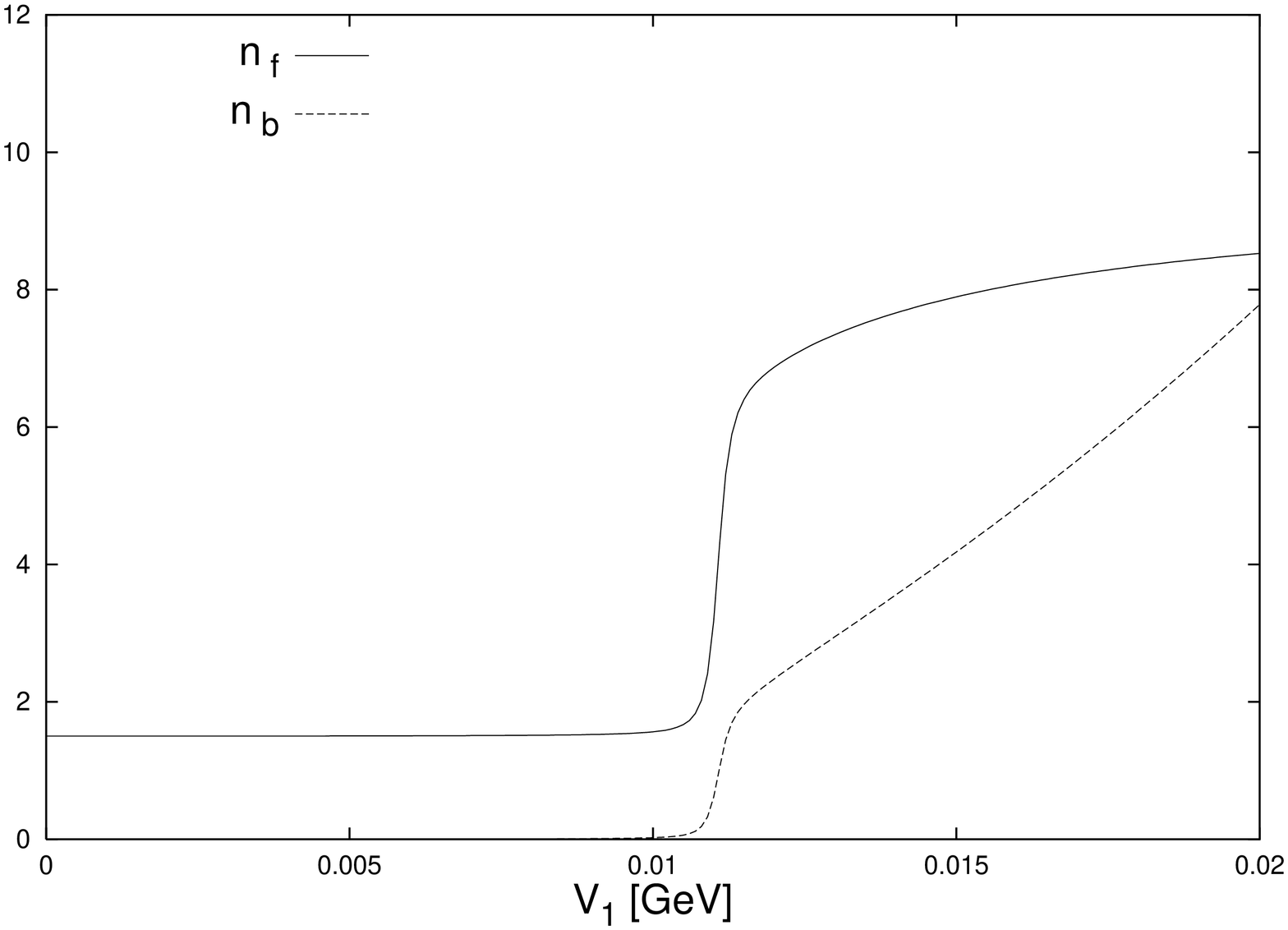,width=10.0cm,height=7cm,angle=270}
\end{picture}

\vskip 1cm \noindent {\bf Figure 7}: The content of
quark-antiquark ($n_f$) and gluon ($n_b$) pairs in the first
excited state of the barionic spectrum, as a function of the
interaction strength $V_1$. The Hamiltonian $\bd{H}_{II}$ was
used.

\newpage
\begin{picture}(100,200)(9,9)
\psfig{file=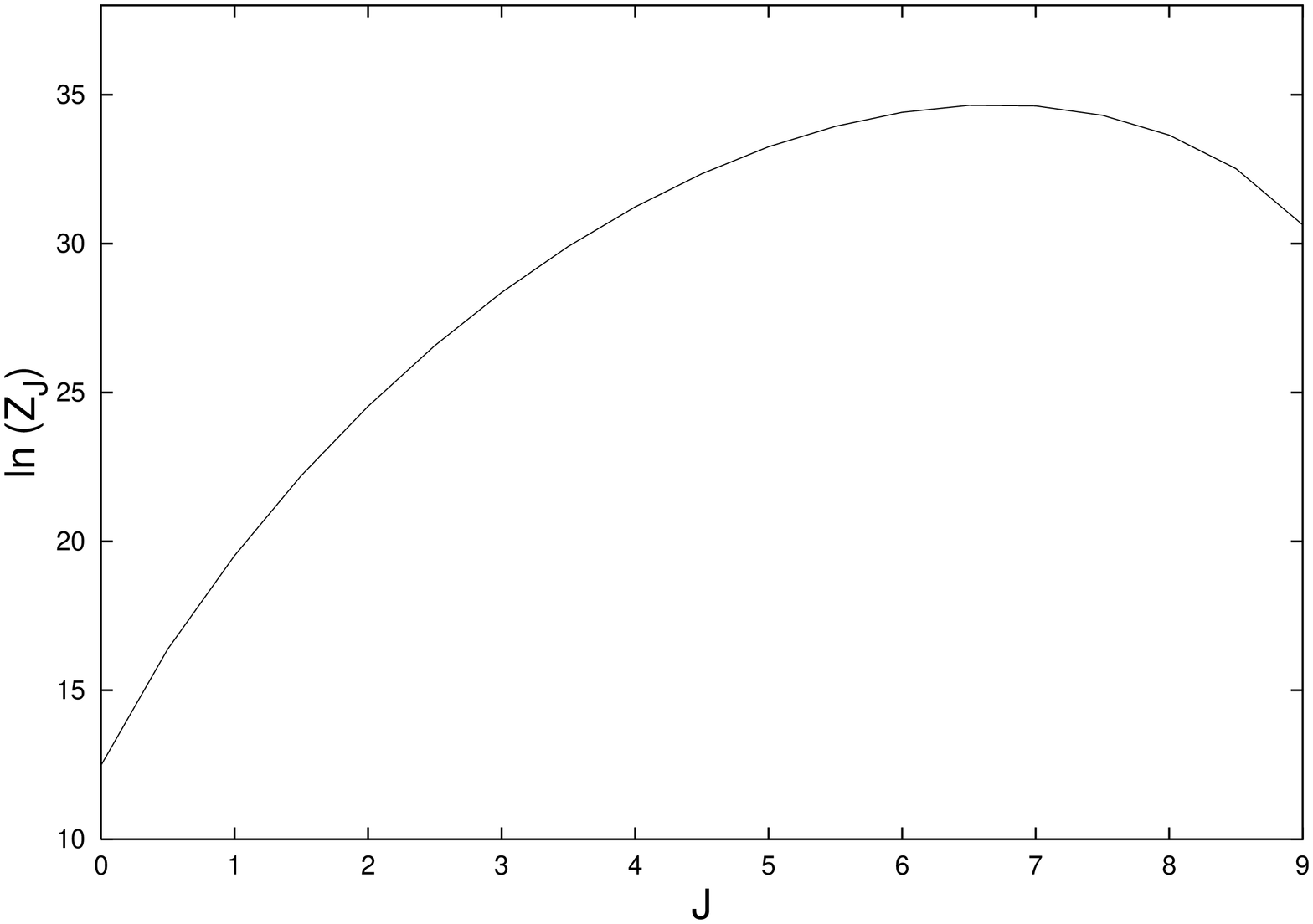,width=10.0cm,height=7cm,angle=270}
\end{picture}

\vskip 1cm \noindent {\bf Figure 8}: Contribution to the partition
function, Eq.(10), for a fixed value of $J$. The curve shows the
logarithm of the results corresponding to the case $\bd{H}_{I}$,
for values of $J\le \Omega=9$, and for $T= 0.2$ GeV and $V_1=0$
GeV.

\newpage
\begin{picture}(100,200)(9,9)
\psfig{file=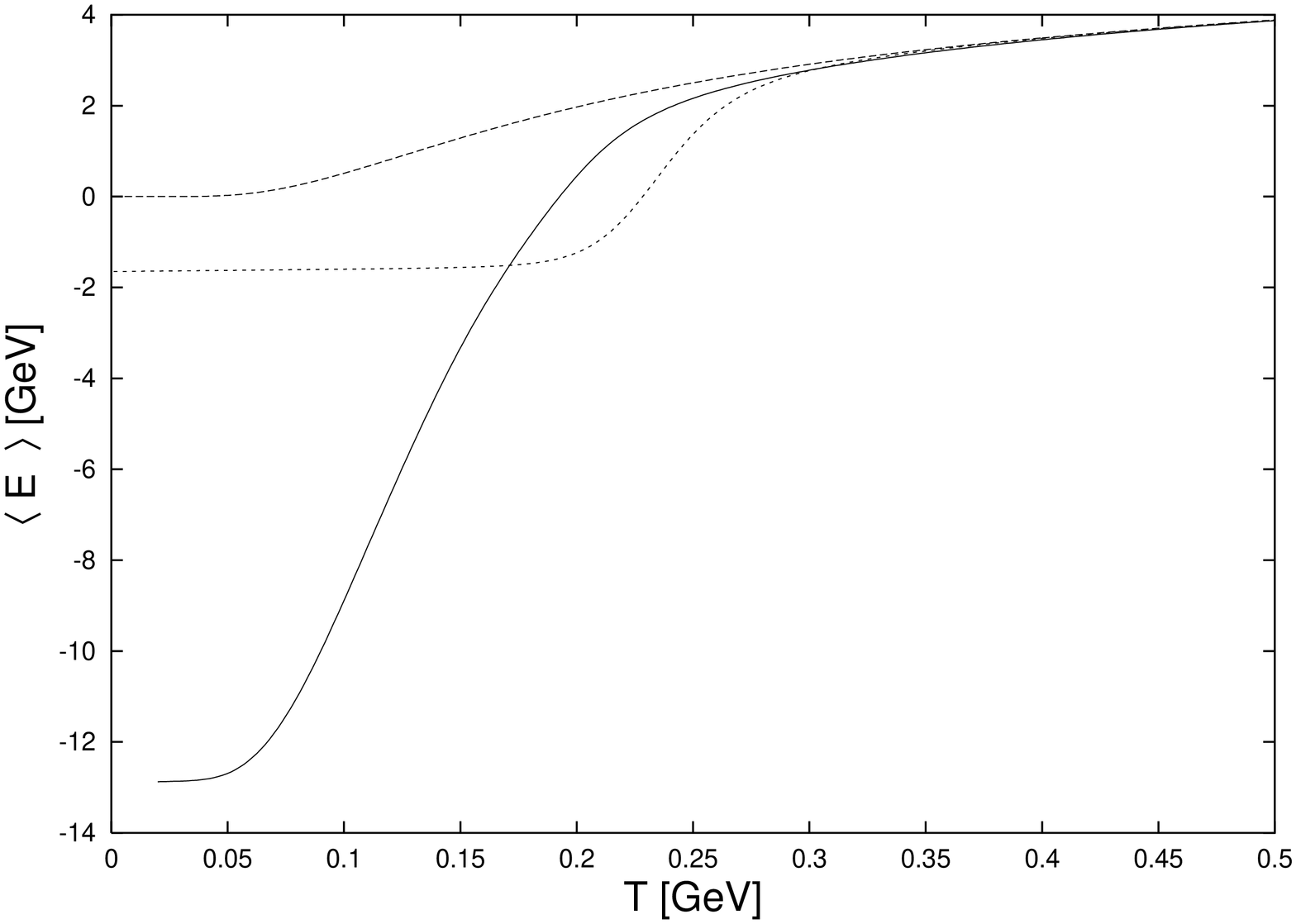,width=10.0cm,height=7cm,angle=270}
\end{picture}

\vskip 1cm \noindent {\bf Figure 9}: The temperature dependence of
the internal energy, as obtained from the calculations performed
with the Hamiltonian $\bd{H}_{I}$. Dashed-lines indicate the exact
results corresponding to the unperturbed ($V_1=0$) case,
small-dashed-lines show the exact results for $V_1=0.04$ GeV, and
solid-line shows the results of the approximations described in
section III of the text, see Eq. (29). The validity of the
approximations (solid-line curve) is limited to $T\ge 0.2$ GeV, as
discussed in the text. The chemical potential has the value
$\mu=0$.

\newpage
\begin{picture}(100,200)(9,9)
\psfig{file=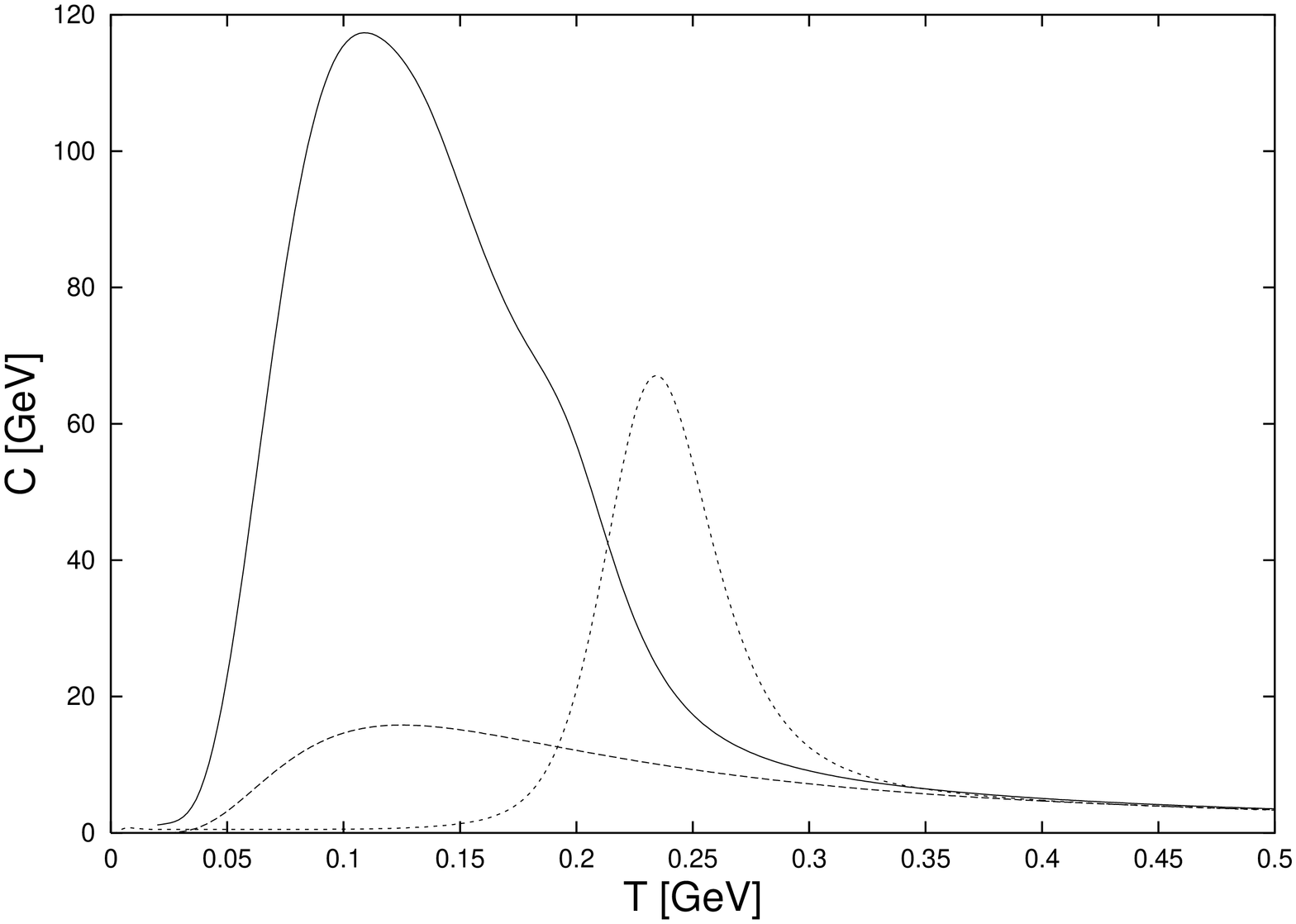,width=10.0cm,height=7cm,angle=270}
\end{picture}

\vskip 1cm \noindent {\bf Figure 10}: The specific heat, as a
function of temperature, for the cases shown in Figure 9. Results
are shown following the notation given in the captions to Figure
9.

\newpage
\begin{picture}(100,200)(9,9)
\psfig{file=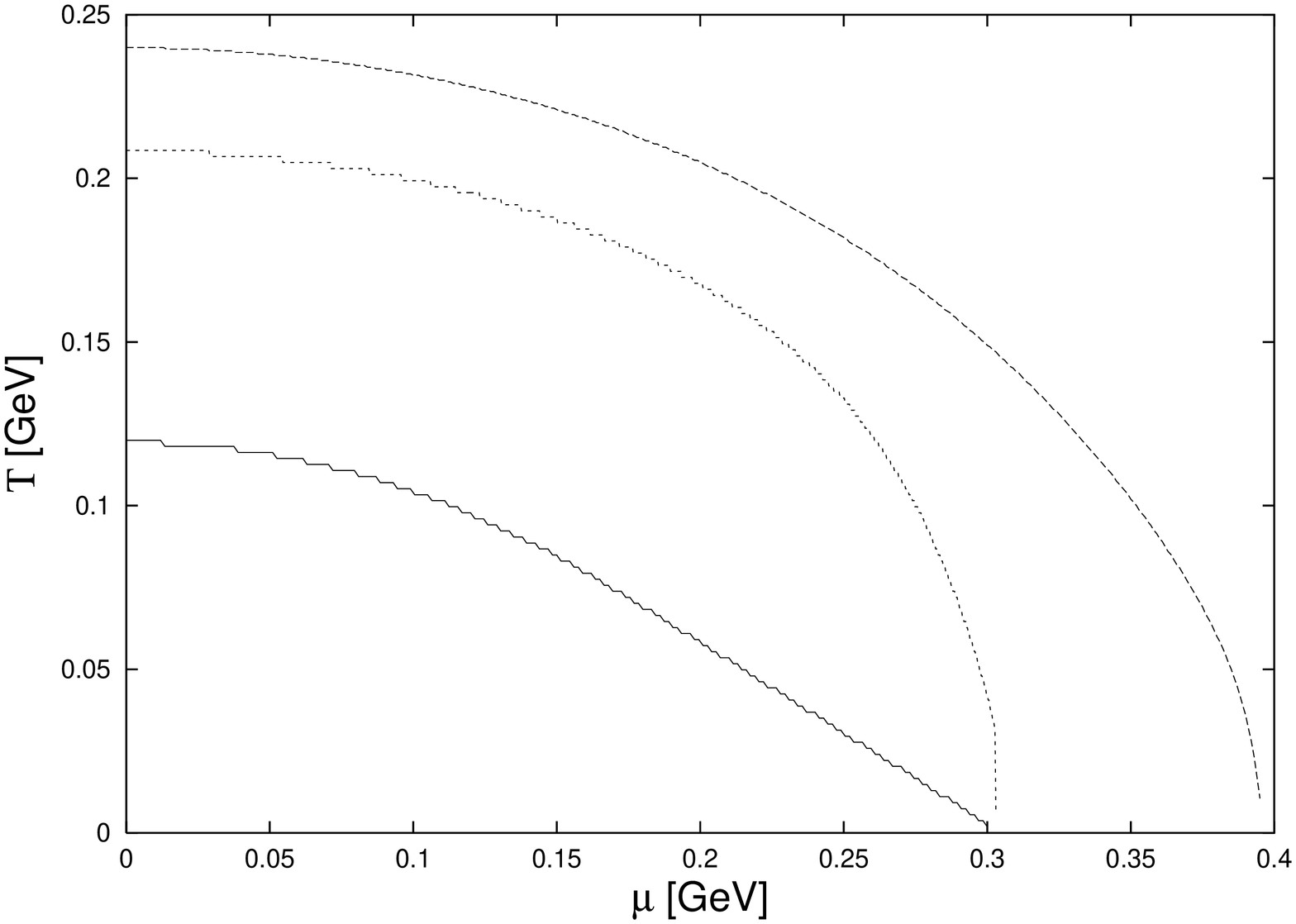,width=10cm,height=7cm,angle=270}
\end{picture}

\vskip 1cm \noindent {\bf Figure 11}: Temperature as a function of
the chemical potential. Solid-line indicates the exact results
corresponding to the unperturbed ($V_1=0$) case, dashed-line shows
the exact results for $V_1=0.04$ GeV, and small-dashed-line shows
the results of the approximations described in section III.

\end{document}